\documentclass[11pt,twoside]{article}

\usepackage{asp2006}
\usepackage{epsf}
\usepackage{psfig}
\usepackage{lscape}

\begin{document}
\title{Uncovering the Active Galactic Nuclei in Low-Ionization Nuclear
  Emission-Line Regions with {\it Spitzer}}   %%% Fill in title
\author{David Rupke, Sylvain Veilleux, DongChan Kim}
\affil{University of Maryland, College Park, MD 20742, USA}
\author{Eckhard Sturm, Alessandra Contursi, Dieter Lutz}
\affil{Max-Planck-Institut f\"ur extraterrestrische Physik, Postfach
  1312, D-85741 Garching, Germany}
\author{Hagai Netzer, Amiel Sternberg, and Dan Maoz}
\affil{Tel Aviv University, Tel Aviv 69978, Israel}

\begin{abstract} The impact of active galactic nuclei on
  low-ionization nuclear emission-line regions (LINERs) remains a
  vigorous field of study.  We present preliminary results from a
  study of the mid-infrared atomic emission lines of LINERs with the
  {\it Spitzer Space Telescope}.  We assess the ubiquity and
  properties of AGN in LINERs using this data.  We discuss what powers
  the mid-infrared emission lines and conclude that the answer depends
  unsurprisingly on the emission line ionization state and, more
  interestingly, on the infrared luminosity.
\end{abstract}

\section{Introduction}

The nature of low-ionization nuclear emission-line regions (LINERs) in
galaxies remains a topic of intense research 25 years after their
discovery.  The {\it Spitzer Space Telescope} has recently made
possible very sensitive observations of the mid-infrared
fine-structure emission lines in these sources.  This gives us new
leverage with which to understand the relative contribution of active
galactic nuclei, shocks, and stellar populations to the energy of
their line emission.

We pose three questions of interest.  (1) Exactly how many LINERs host
an AGN?  This question is still open.  (2) Does the AGN excite the
low-ionization mid-infrared emission lines?  This has been an active
research question in the optical band, but we re-pose it for
mid-infrared observations.  (3) How do infrared-luminous and
infrared-faint LINERs differ in their mid-infrared spectra?  Beyond
what was previously known from other wavelengths, \citet{sturm06a}
show that these sources differ greatly in their mid-infrared spectra,
including spectral energy distributions, dust emission features, and
atomic emission lines.  Here we address in more detail the differences
in atomic emission lines.

We present ongoing results from our {\it Spitzer} study using the {\it
  Infrared Spectrograph}.  We observed 33 galactic nuclei at
wavelengths from $5-37~\mu$m and resolving powers of $\sim$100
($5-15$~$\micron$) and $\sim$600 ($10-37$~$\micron$).  Our two samples
of infrared-luminous and infrared-faint LINERs are from \citet{kim95a}
and \citet*{hfs97a}, respectively.  Previous results from this study
can be found in \citet{sturm05a} and \citet{sturm06a}.  The full
details, including data reduction, measured line fluxes, and
photoionization modeling, are in preparation (Rupke et al. 2007, in
prep.).

\section{High-Ionization Emission Lines}

There are three mid-infrared transitions that are common in our
spectra and arise from ions with high enough ionization potentials to
be good AGN signatures: [O~{\footnotesize IV}] 25.9~$\micron$ and
[Ne~{\footnotesize V}] 14.3~$\micron$ and 24.3~$\micron$.  The
[Ne~{\footnotesize V}] transitions are faint in our sources, and we
detect the 14.3~$\micron$ line in 40\% of IR LINERs and 0\% of
IR-faint LINERs.  The [O~{\footnotesize IV}] transition is brighter
and is visible in 90\% of both samples.  The latter can have some
contribution from star forming regions, however.

To address this possible mixing, we show in Figure 1 that there is
significant excess [O~{\footnotesize IV}] emission in LINERs compared
to starburst galaxies of a given 60~$\micron$ luminosity.  We
interpret this as a signature of a non-stellar power source for most
of the [O~{\footnotesize IV}] emission in LINERs: probably an AGN.
However, there is less [O~{\footnotesize IV}] emission than in Seyfert
galaxies of the same luminosity, giving credence to the
``low-luminosity AGN'' scenario.

\begin{figure}[!ht]
\plottwo{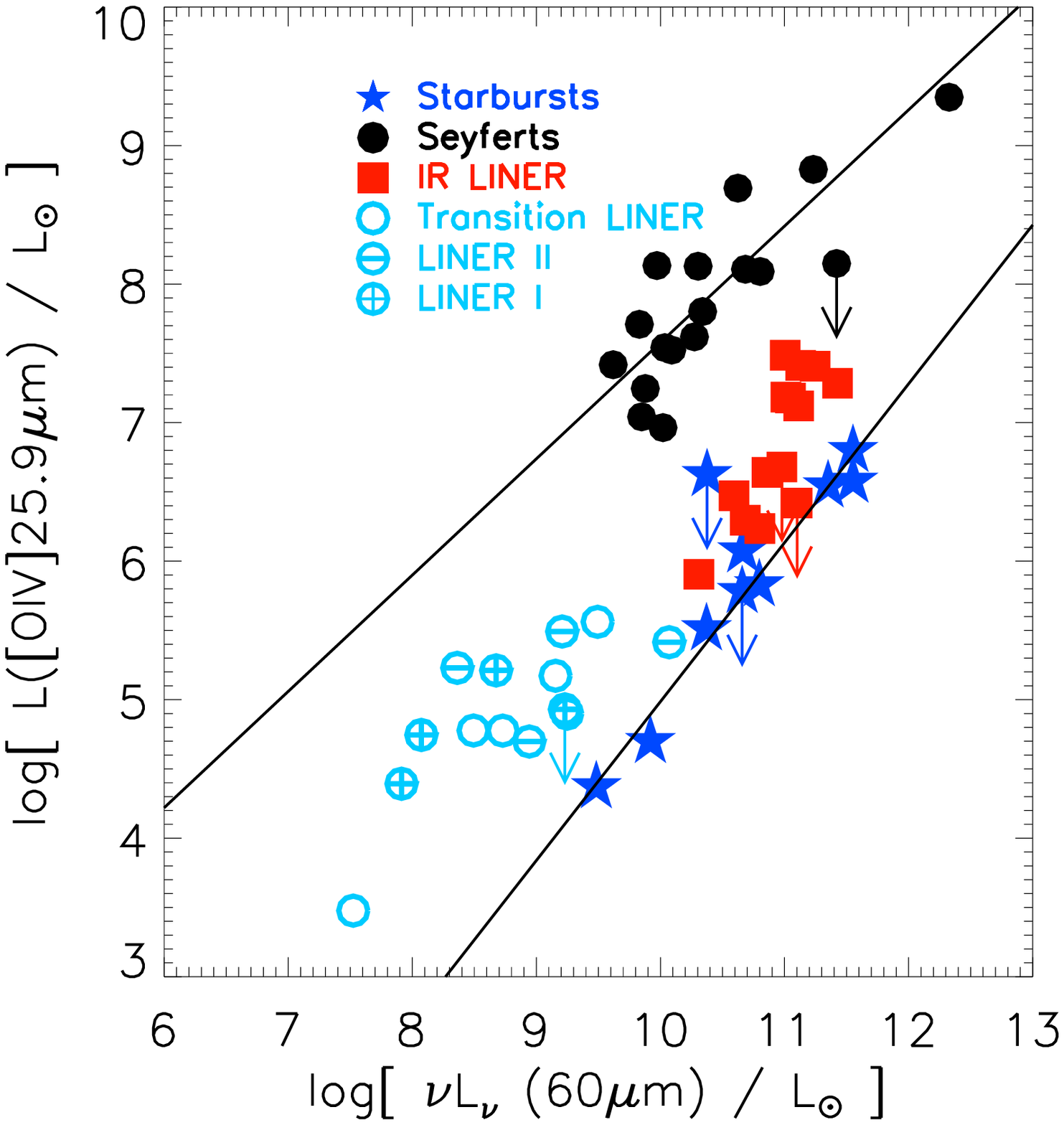}{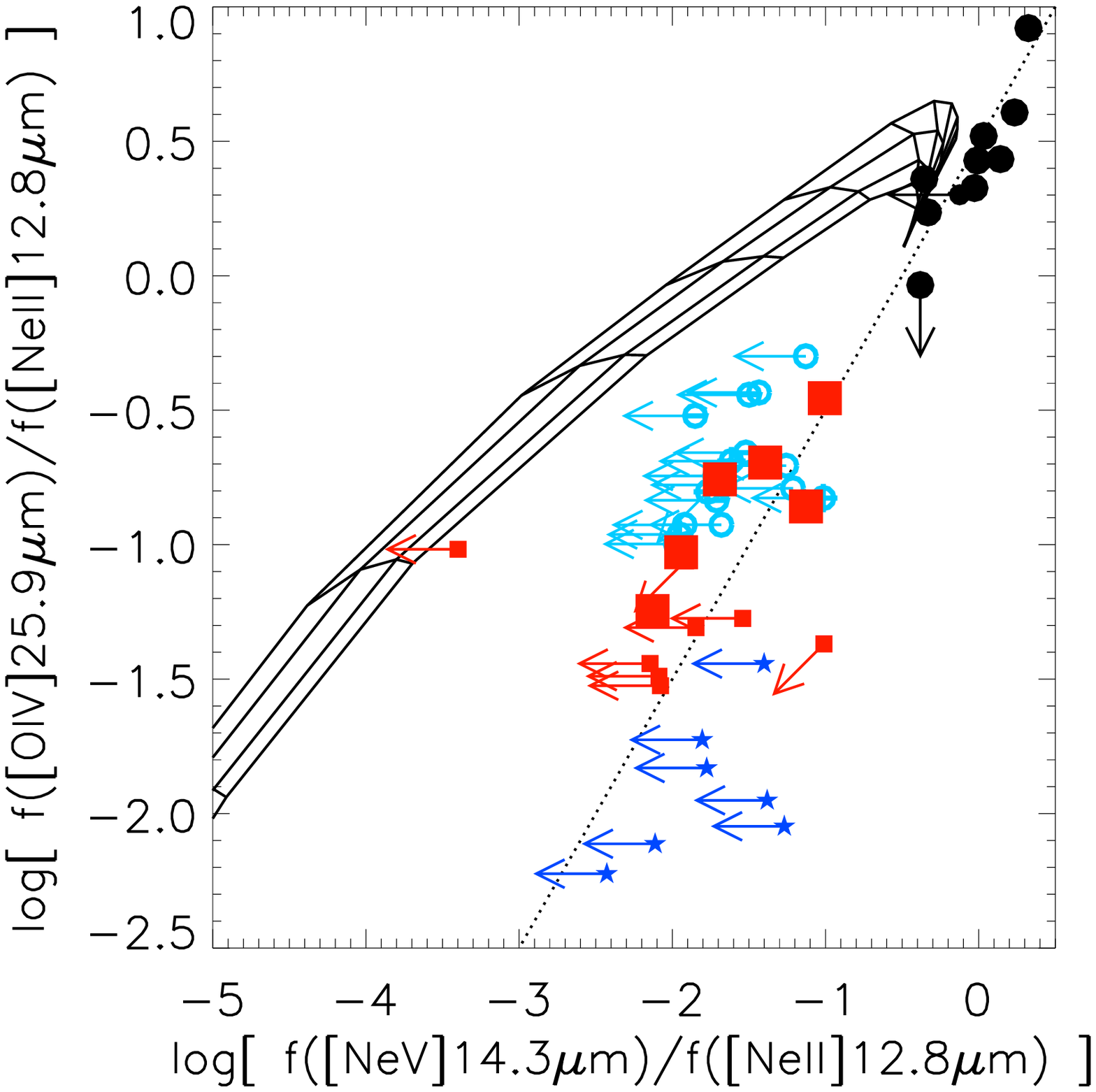}
\caption{{\it(left)} [O~{\scriptsize IV}] 25.9~$\micron$ luminosity
  vs. 60~$\micron$ continuum luminosity, for starbursts and Seyferts
  (stars and filled circles; \citealt{sturm02a,verma03a}) and our
  LINERs (squares for IR LINERs, open circles for IR-faint LINERs).
  The lines are separate linear fits to the starbursts and Seyferts.
  The three IR-faint LINER categories are discussed in \citet{hfs97a}.
  {\it(right)} A line ratio diagram of high-ionization lines
  normalized to an H~{\scriptsize II}-region line, [Ne~{\scriptsize
    II}] 12.8~$\micron$.  The straight line is a starburst mixing
  line; starburst contribution increases down and to the left.  The
  curved lines are AGN photoionization models from \citet*{gds04a},
  where ionization parameter decreases down and to the left.}
\end{figure}

LINERs also have excess [O~{\footnotesize IV}] emission compared to
starbursts when normalized by [Ne~{\scriptsize II}] 12.8~$\micron$, an
H~{\footnotesize II}-region tracer (Figure 1).  We compare observed
line ratios to AGN photoionization models from \citet*{gds04a}, and
assume that the starburst contribution affects primarily the
[Ne~{\scriptsize II}] line.  The conclusion is that LINERs are
consistent with either (1)~AGN of moderate or Seyfert-like ionization
parameter mixed with a starburst or (2)~AGN with low ionization
parameter.  IR LINERs are on average more like the former, while
IR-faint LINERs are inconsistent with Seyfert-like ionization
parameter.

\section{Low-Ionization Emission Lines}

LINERs are defined by the relative strength of their optical
low-ionization emission lines.  It turns out that certain
low-ionization lines are strong in the mid-infrared, as well, compared
to starbursts.  The [Fe~{\footnotesize II}] 26.0~$\micron$ and
[Si~{\footnotesize II}] 34.8~$\micron$ lines are both stronger than in
starbursts (relative to [Ne~{\footnotesize II}]).  This is especially
true of [Fe~{\footnotesize II}], whose relative strength is greater
than in starbursts by 50\% in IR LINERs and a factor $\sim$10 in
IR-faint LINERs (Figure 2).

Fe and Si are both heavily depleted in the ISM, but the lack of strong
[Ca~{\footnotesize II}] emission at 7291~\AA\ and 7324~\AA\ in LINERs
\citep*[e.g.,][]{hfs93a} argues against this being due to grain
destruction \citep{vb96a}.  Alternatively, emission from these two
observed transitions can arise from photo-dissociation and X-ray
dissociation regions (PDRs and XDRs).  The flux ratio of the two lines
can thus constrain PDR and XDR models \citep{kwh06a,msi06a}.

In Figure 2 we show that there is a significant difference in this
ratio between starbursts and LINERs, especially for IR-faint LINERs.
There is also a difference between IR-luminous and IR-faint LINERs.
The sense of the difference is such that, compared to IR LINERs,
IR-faint LINERs should on average have a higher far-UV flux incident
on the gas in PDR models and a higher X-ray flux in XDR models (M.
Wolfire, pvt.  comm.; R. Meijerink, pvt. comm.).  This is contrary to
the observation of intense star formation in IR LINERs (which is
accompanied by strong UV flux) as well as the absence of a strong
correlation between X-ray and IR fluxes in LINERs \citep{dudik05a}.

\begin{figure}[!ht]
\plottwo{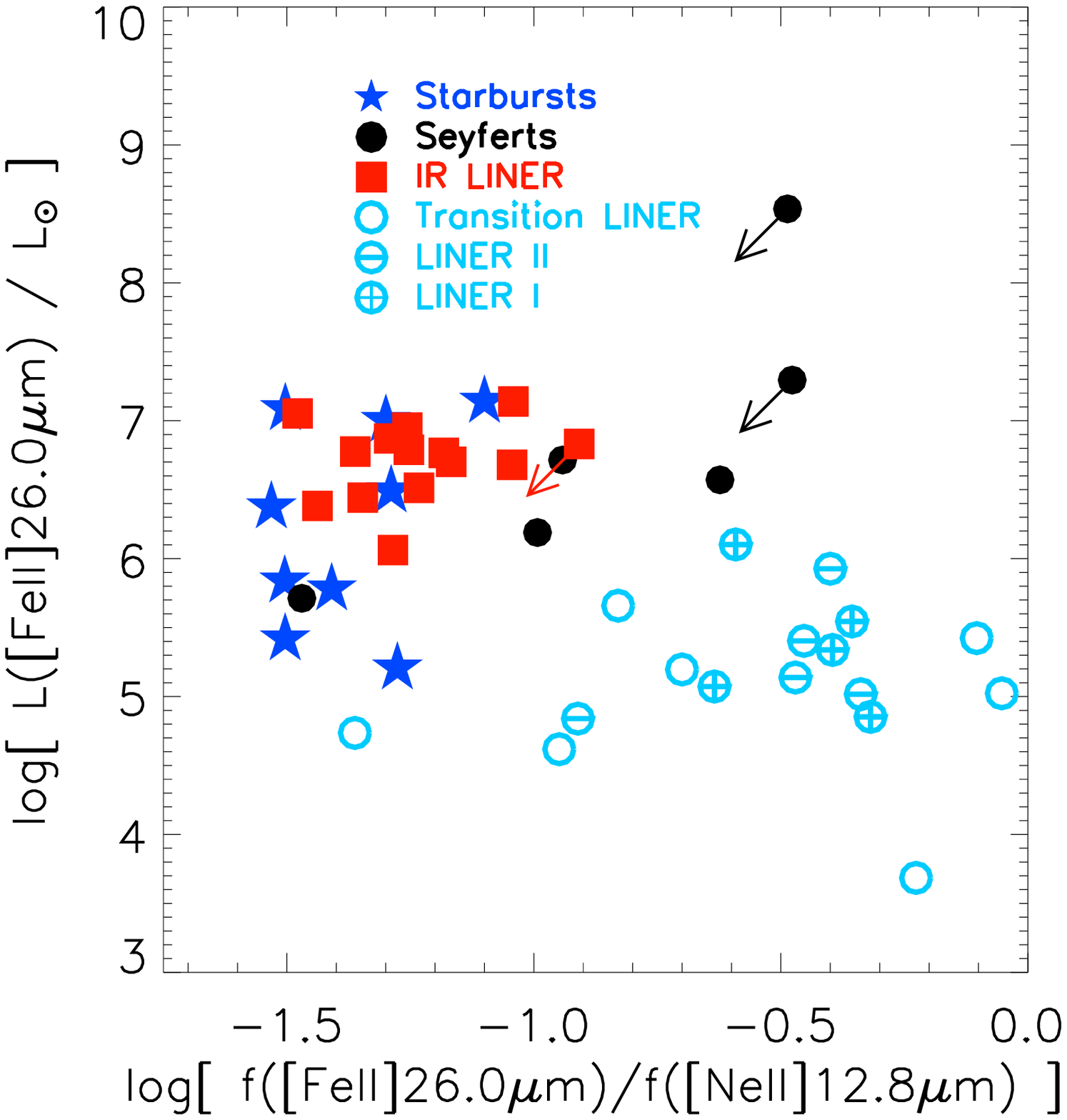}{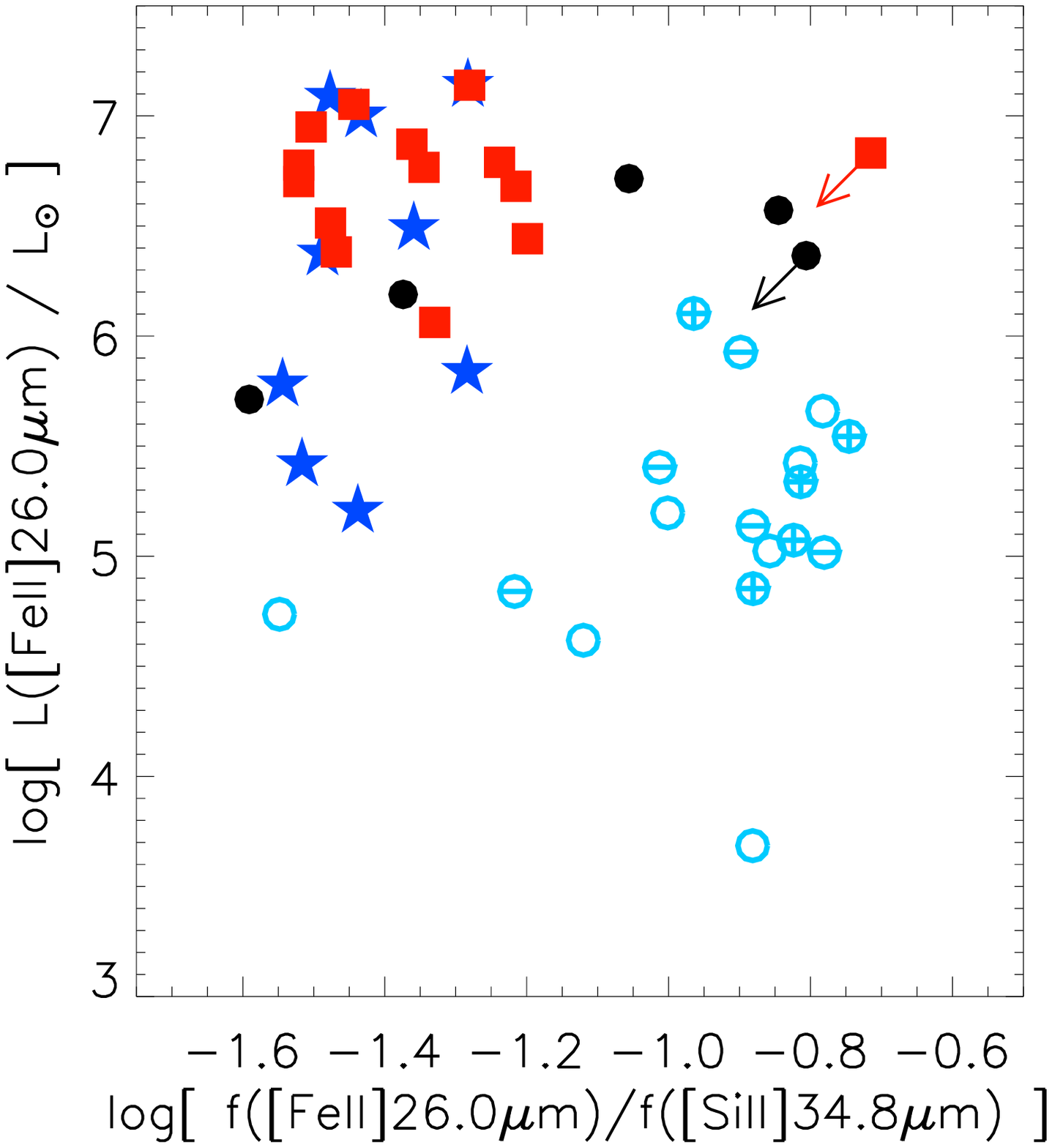}
\caption{Two diagnostic line ratios vs. the [Fe~{\scriptsize II}]
  26.0~$\micron$ line luminosity.  LINERs, especially IR-faint ones,
  have excess [Fe~{\scriptsize II}] {\it (left)} and [Si~{\scriptsize
    II}] 34.8~$\micron$ (not shown) emission compared to starbursts
  when normalized by [Ne~{\scriptsize II}].  {\it (right)} Because
  both [Fe~{\scriptsize II}] and [Si~{\scriptsize II}] are strong in
  PDRs and XDRs, the ratio of the two provides information on a
  region's ionizing flux and density.}
\end{figure}

Since the same model is not applicable to both subsamples
simultaneously, we note that PDRs are more natural in the case of IR
LINERs because of the strong star formation present.  We thus conclude
that the [Fe~{\footnotesize II}] 26.0~$\micron$ and [Si~{\footnotesize
  II}] 34.8~$\micron$ lines may be powered by PDRs in IR LINERs and
XDRs in IR-faint LINERs.  However, this conclusion is preliminary and
more work is needed to confirm both the applicability of these models
and the conclusions drawn from them.

\section{Summary}

We now revisit the three questions we posed earlier.  (1) Exactly how
many LINERs host an AGN?  We have shown that the mid-infrared
high-ionization lines are consistent with a very high AGN fraction.
(2) Does the AGN excite the low-ionization mid-infrared emission
lines?  As in the optical, the LINERs have excess mid-infrared
low-ionization line emission, especially in lines that are tracers of
PDRs and XDRs.  The IR-faint LINERs are good XDR candidates, and as
such may indicate the presence of an AGN.  (3) How do
infrared-luminous and infrared-faint LINERs differ in their
mid-infrared spectra?  We have shown that the mid-infrared atomic
emission-line data are consistent with IR-luminous LINERs having
different high- and low-ionization line properties than IR-faint
LINERs.  Our analysis suggests that, on average, IR-luminous LINERs
have buried AGN with moderate-to-high ionization parameter and may
contain strong PDR emission, while IR-faint LINERs have lower
ionization parameter AGN and possibly strong XDR emission.

\acknowledgements

We thank Brent Groves for useful discussions.  This work is based on
observations made with the {\it Spitzer Space Telescope}, which is
operated by JPL/Caltech under NASA contract 1407. Support for this
work was provided by NASA through contracts 1263752 and 1267948 issued
by JPL/Caltech.

\end{document}